\documentclass[a4paper,12pt]{elsart}
\usepackage{epsfig}
\begin{document}
\begin{frontmatter}
\title{Interfacial Properties of Nonionic Micellar Agregates
        as a Function of Temperatures and Concentrations}

\author{Luca Falconi, Marco Maccarini, Giuseppe Briganti}
\address{Department of Physics,
         Univiversity ``La Sapienza'',
         P. A. Moro, 2 00185 Rome, Italy}
\author{Giovanni D'Arrigo}
\address{Department of Energetic,
         Univiversity ``La Sapienza'',
         P. A. Moro, 2 00185 Rome, Italy}

\begin{abstract}
By  means  of density, dielectric spectroscopy  and  sound
velocity measurements we perform a systematic study on the
polyoxyethylene    $C_{12}E_{6}$    nonionic    surfactant
solutions  as a function of temperature and concentration.
Both   density   and  sound  velocity   data,   at   about
$34^{\circ}C$, coincide with the value obtained  for  pure
water.  Above this temperature the density is  lower  than
the  water  density  whereas  below  it  is  greater,  the
opposite   happens  for  the  compressibility.   Combining
results  from these different techniques we tempt  a  very
detailed  description  of the evolution  of  the  micellar
interfacial  properties with temperature. It  is  well
known   that   nonionic   surfactant   solutions dehydrate,
growing temperature. Our results  indicate  that  this
process  is  associated with a continuous  change  in  the
polymer  conformation  and in the  local  density  of  the
micellar interface.
\end{abstract}
\end{frontmatter}

\section{Introduction}
The thermodynamic properties of the polyoxyethylene nonionic amphiphile 
($C_iE_j: C_iH_{2i+1}-[O-CH_2-CH_2]_j OH$) water mixtures are
characterised by a very delicate balance between competitive forces;
small temperature variations can change the overall equilibrium condition
of the solution \cite{cork}. These mixtures present a temperature
dependence of the aggregate dimension \cite{Lind}, a lower
consolutum temperature and a very flat coexisting region, various
liquid crystal phases \cite{mitc} . These properties often resemble
each other for the different $i, j$ monomeric species using specific
rescaling parameters (\cite{Lind}, \cite{puvv}).
Changes in the hydration state of the micellar aggregates are
considered the driving mechanism for micellar growth \cite{puvv}
and phase separation, both in these surfactant solutions \cite{kje}
as well as in pure $EO_j(O-CH_2-CH_2)_j$-water mixtures \cite{kje2}.
Clearly the
volume per monomer is directly related to the degree of hydration,
i.e. different monomeric volumes stabilise different micellar shape
and size distribution function. Since the oil core cannot easily change
its density and it is hydrophobic, the variations of the degree
of hydration and then of the monomeric volume are
mainly due to the properties of the $EO$-water
mixtures at the micellar interface. This point of view naturally
divides the solution into three subregions: the oil core,
with overall properties similar to an oil bulk phase,
the interfacial EO-water mixture and the bulk water phase.
The three regions are weighted by the relative volume fraction.

Experimental techniques which can contrast the three different
regions of the amphiphilic solutions are useful to determine
the thermodynamic properties of the interface as a function of
temperature and surfactant concentration. To this end
we utilised density measurements (oil bulk phase $\sim 0.75 g/cm^3$;
water bulk phase $\sim 0.99 g/cm^3$; EO-water mixture ranging
from $0.998 g/cm^3$ and $1.10 g/cm^3$ at $25^{\circ}C$); sound
velocity (oil bulk phase $\sim 1297m/s$ ; water bulk phase $\sim 1450m/s$;
EO-water mixture
ranging from $1500 m/s$ and $1600 m/s$ at $25^{\circ}C$) and dielectric constant data
(oil bulk phase $\sim 2$; water bulk phase $\sim 80$; EO-water mixture
ranging from $30$ and $60$ at $25^{\circ}C$). All
these techniques are linear in the volume fraction of the relative three 
components, but related to different properties of EO-water phase.
The density depends on the mass concentration at the interface;
the dielectric costant on the local polarizability, related to the
molecular connectivity of the EO-water solution; the sound velocity is
related to the elastic properties of the interfacial mixture.


\section{Material and Methods}
The sample analyzed ($C_{12}E_6$) was purchased from Nikko Chemicals
and was diluted without further purification. The concentration of the
different water solutions were stated by weighing.
Density measurements were performed with PAAR Digital Density Meter DMA60
in combination with remote cell DMA 602, that can provide an accuracy
up to $\pm 1.5 \cdot 10^{-6} g/cm^3$, according to the oscillating
sample tube method. The external cell was thermostated by Heto DBT 6
Thermostat and Heto CB8-30e Cooling Bath that guarantees a temperature
stability within $0.05^{\circ}C$.  
A variable path-length cell was used for the ultrasonic measurements. The cell 
content is about $30 ml$ of solution. Two matched 5 MHz fundamental crystals were 
used, along with a quartz delay line, in a pulsed-sound mode. The temperature of the 
sample was controlled by a water thermostat regulated within $\pm 0.1K$. The sound 
velocity (c) was obtained by acoustically overlapping pulses under variable-path 
conditions to coherent CW signals generated by a Matec Pulse Modulator equipment 
\cite{arri}. The measurements were performed, at 5 and 45 MHz, the accuracy of the 
absolute sound velocity data (at fixed frequency) is estimated to be $\pm 0.2 m/s$
over the whole frequency range.


\section{Data  Analysis}
Some interesting indications are obtained from the density of the pure
liquid surfactant. The results for $C_{12}E_5$, $C_{12}E_6$,
$C_{12}E_7$ and $C_{12}E_8$ are
\begin{figure}
\begin{center}
\epsfig{file=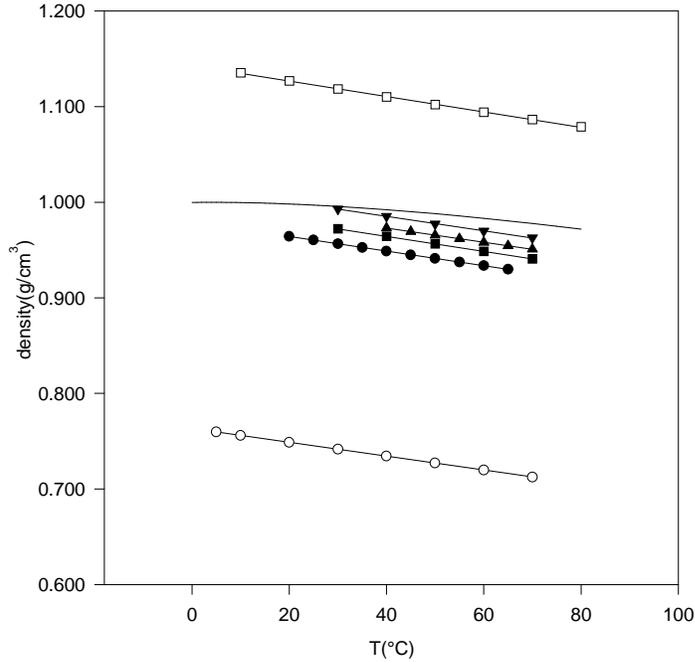,height=10cm}
\caption{
The  density  of  EO  polymer  in  its  liquid  state,  of pure
$C_{12}E_5$ (filled circle),  $C_{12}E_6$ (filled square),
$C_{12}E_7$ (filled triangle up),  $C_{12}E_8$ (filled triangle down)
liquid surfactant, of water ( full line), of EO (open square)
 and of dodecane  (open circle) are reported.}
\label{fig:dejp}
\end{center}
\end{figure}
reported in figure \ref{fig:dejp}, for comparison the pure PEO, the water and the bulk
dodecane densities are reported too. The observed experimental values
for the different $C_{12}E_j$ are between the oil and the PEO density,
$C_{12}E_8$
is closer to the PEO density whereas $C_{12}E_5$ has a lower density, closer
to the data of the oil phase. The lines trough the data represent best
fit obtained combining the density of the bulk phases, namely dodecane
and EO polymers \cite{messina}, weighted with the relative volume fraction:
\begin{equation}   \label{eq:rhopu}
\rho = \phi_h \rho_h + \phi_t \rho_t 
\end{equation}
The fitting parameters, namely the density of the EO phase and the one
of the oil core, practically coincide with the experimental one
observed in the two bulk phases. Besides, if at the density of $C_{12}E_5$
we add ideally the contribution of n EO unit, i.e. in eq. \ref{eq:rhopu}
we change only
the volume fraction since we add to the total volume and to the head
group volume the contribution of n EO units, the experimental density of
the $C_{12}E_{5+n}$ is obtained.

\begin{figure}
\begin{center}
\epsfig{file=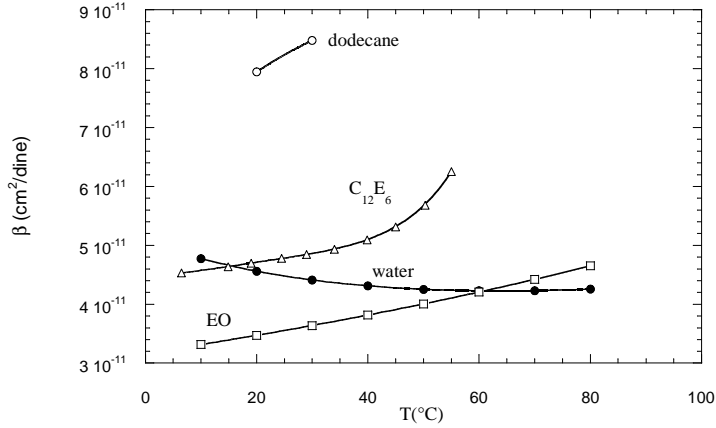,height=10cm}
\caption{
The compressibility the pure $C_{12}E_6$ surfactant
in  its liquid state is reported and compare with the  one
of dodecane, EO polymer and water.}
\label{fig:betapurivst}
\end{center}
\end{figure}

Similar analysis can be applied on the adiabatic compressibility
defined as \cite{librotermo}:
\begin{equation}
 \beta = \frac{1}{\rho c^2} = \frac{1}{\rho}
         \bigg[ \frac{\partial \rho}{\partial P} \bigg]  =
                              \frac{-1}{V}
         \bigg[ \frac{\partial V}{\partial P} \bigg]
\end{equation}
where $\rho$ is the density, $c$ the sound velocity, $P$ the pressure and $V$
the volume.
In Fig. \ref{fig:betapurivst} are presented the results from pure $C_{12}E_6$
with the one of the EO polymer and of the dodecane.
In this case too the compressibility of the pure surfactant can be
written as a linear combination of the one of the polymeric phase
and one of the oil phase; computing the derivative of eq. \ref{eq:rhopu}
the adiabatic compressibility of the pure surfactant results:
\begin{equation}
 \beta \underline{V} = \beta_t \underline{V}_t - x_h \beta_h \underline{V}_h 
\end{equation}
where the undersigned volume are the molar volume of the pure surfactant, of the 
oil phase and of the EO phase respectively. The fitting parameters give values with 
few percent variation from the one of the relative bulk phases of the EO polymer and 
dodecane ($\beta_h$ and $\beta_t$).
These results indicate that even at $100\%$ surfactant concentration still two segregate 
phases are present: the oil and the polymer. The oil phase is hydrophobic then it will 
not change composition by adding water. The EO-water interfacial region match the 
interaction of the oil phase with the bulk water, minimizing the osmotic contribution. 
Then in the analysis of the experimental results on the
$C_{12}E_6$-water solutions, we
consider the oil phases equivalent to a bulk dodecane phase at the same temperature. 
For what concerns the water phase we can use the bulk water properties at the same 
temperature if the $C_{12}E_6$-water solutions are diluted.
Under this basis we attribute the
variation of the properties of the solution to variation of the EO-water phase, for 
examples EO conformation, degree of hydration etc.

\begin{figure}
\begin{center}
\epsfig{file=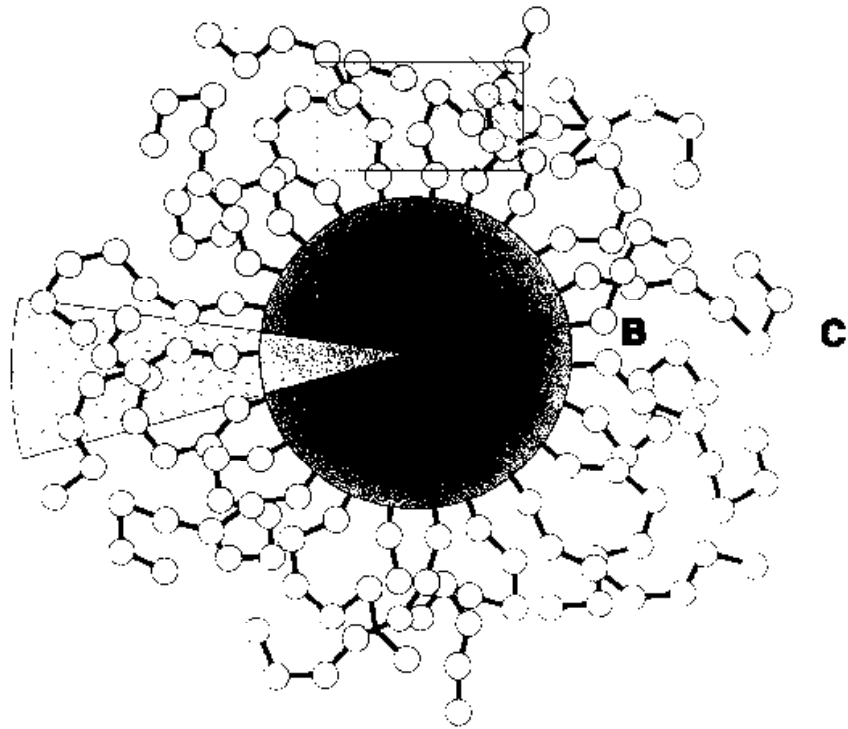,height=10cm}
\caption{
Schematic view of a cross section of  a  micellar
aggregate  (spherical or cylindrical). The  area  named  A
represents  the oil core region, B the interfacial  region
with EO termination and water, C the bulk water phase. The
volume  indicated on the left side represents the hydrated
volume  per monomer, divided in its part: head  and  tail)
the  rectangle  at the top schematise the interfacial  EO-
water mixture.}
\label{fig:modmic}
\end{center}
\end{figure}

In Fig. \ref{fig:modmic}  a schematic view of a cross section of a spherical
or cylindrical micelle is reported.
In our model of nonionic micellar solutions the
oil core volume (A) is obtained from the bulk oil phase density. The
partial molar volume of the bulk water (C) is known. On the other hand
the interfacial volume of the head group termination (B) can greatly
vary. Depending on the polymer conformation, the polymer molarity in the
B region change and different ammount of water are needed to fill it up
to the right density.
All our experimental quantities depend linearly on the hydrated volume
per monomer $V_h$ of the polymeric terminations at the micellar interface.
No univocal indication can be obtained about this volume from other
experiments. Then in our analysis we extract from the experimental
results of $C_{12}E_6$-water solutions the interfacial density, adiabatic
compressibility and dielectric constant as a function of the parameter
$V_h$ and of the temperature.
In particular for what concerns density we obtain the interfacial values
starting from an equation for the solution density given by:
   \begin{equation}       \label{eq:rho3}
    \rho= \phi_w \rho_w  +  \phi_s \rho_s     \; .
   \end{equation}
where $\rho_w$, $\phi_w$ and $\rho_s$, $\phi_s$ indicate
the density and the volume fraction rispectively of the water and the
surfactants. $\rho_s$ represents the density of the micelle and, if
we define
 $\rho_h$, $\phi_h$ and $\rho_t$, $\phi_t$
the density and the volume fraction respectively of the interface
and the oil core, we can write
   \begin{equation}       \label{eq:rhos}
    \rho_s= \phi_t \rho_t  +  \phi_h \rho_h     \; .
   \end{equation}
Defining $M_w$, $V_w$ and  $M_s$, $V_s$ the mass and the volume of
respectively a water and a surfactant molecule, $N_s$ the number of
surfactant molecules and $N_{w}^{h}$  the number of water molecule
per polar head in the interface, then by definition we have:

    \begin{equation}           
     \rho_w = \frac{M_w}{V_w}  \qquad \qquad
             \rho_s = \frac{M_s+N_{w}^{h}M_w}{V_s}
    \end{equation}
    \begin{equation}
         \phi_w = \frac{(N_w-N_{w}^{h}N_s)V_w}{V} \qquad \quad
         \phi_s = \frac{N_s V_s}{V}    \;.
    \end{equation}

Using these equations $\rho_h$ is given by:
   \begin{equation}  \label{eq:rhoh}
    \rho_h= \rho+ \frac{1}{V_h} \bigg[ V_t(\rho -\rho_t)   +
           \frac{N_w-N_{w}^{h}N_s}{N_s}V_w (\rho -\rho_w)    \bigg] \;.
   \end{equation}
To eliminate the dependence from the unknown $N^{h}_{w}$ variable,
we use the condition for the interfacial density:
  \begin{equation}
   \rho_h= \frac{(PM)_h+ N_{w}^{h} (PM)_w}{N_A V_h} \; ,
  \end{equation}
that  finally give the following equation:
    \begin{equation}                   \label{eq:rhoh2}
     \rho_h= \frac{\rho_w}{\rho} \Bigg{\{}
                \rho+ \frac{1}{V_h} \bigg[ V_t(\rho -\rho_t)   +
            \bigg(  \frac{N_w}{N_s}+ \frac{(PM)_h}{(PM)_w}    \bigg)
                            V_w (\rho -\rho_w)
                                                      \bigg] \Bigg{\}}
    \end{equation}
The last equation represents the density $\rho_h$ of the micellar interface
parametrized on its volume $V_h$.

For the adiabatic compressibility  the previous model gives an equation
\begin{equation}           \label{eq:comprh}
 \beta \underline{V} = \beta_w \underline{V}_w
        \big[ 1-x_s (1+ N^{h}_{w})  \big] +   x_s \big[
        \beta_t \underline{V}_t + \beta_h \underline{V}_h \big]
\end{equation}
where $x_s$ is the mole fraction of the surfactant in solution,
$\beta_i$ are the adiabatic compressibilities and the $\underline{V}_i$
the relative molar volumes of the three components in the solution.
$N^{w}_{h}$, as before, represents the number of hydrated water molecules.

The iterative use of of mixtures equation for dielectric costant,
first on the surfactant-water solutions, then on the
micella treated as a two component solution (the oil phase and the EO-water phase), 
gives an expression for the interfacial dielectric constant that parametrically depend 
on the volume $V_h$ of the head region (see \cite{brig} for details).

\section{Experimental results}

\begin{figure}
\begin{center}
\epsfig{file=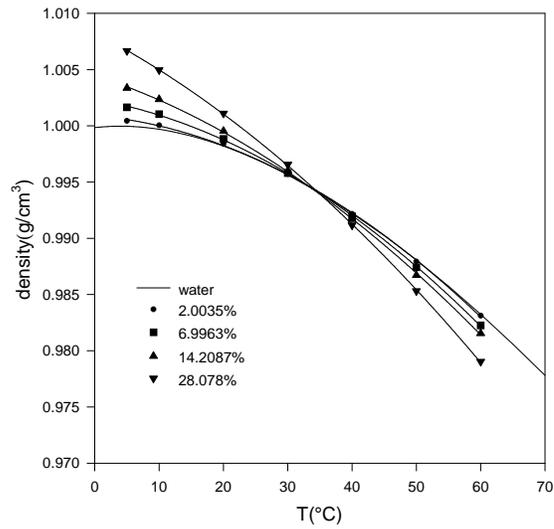,height=8cm}
\caption{
The density of surfactant solutions from 2  to  28
$w\%$  are reported as a function of the temperature,  for
comparison  the water density (continuous  line)  and  the
pure surfactant density are reported too.}
\label{fig:de6}
\end{center}
\end{figure}

The density measurements are reported in Fig. \ref{fig:de6} as
a function  of
temperature  for different concentrations, from 2  to  28  weight$\%$
($w\%$).  For  comparison the water density and the  pure  surfactant
density are reported too as a continuous lines. At low temperature
the  solution densities are higher than the water density  whereas
at high temperature are lower for all the measured concentrations.
The crossing temperature is at about $34^{\circ}C$.
 Similar  results  are observed with the adiabatic compressibility
\begin{figure}
\begin{center}
\epsfig{file=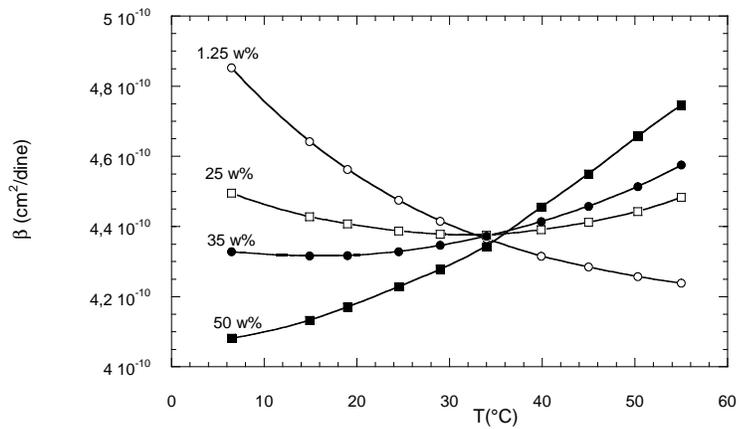,height=10cm}
\caption{
The compressibility of water surfactant solutions
at  different concentration (from 1.25 to 50 $w\%$)  as  a
function of temperature are reported.}
\label{fig:betasolutvst}
\end{center}
\end{figure}
(see  Fig. \ref{fig:betasolutvst}), there exist a crossing temperature, again at  about
$34^{\circ}C$,  where the solution compressibility collapse  at  the  same
values for all the investigated concentrations (from 2 to 50  $w\%$).
It is interesting to notice that there are no ideal combination of
pure  surfactant  and pure water density or compressibility  which
can  give  the experimental $C_{12}E_6$-water values. It means that  the
interfacial  region,  composed  by  EO  terminations   and   water
molecules,  strongly  modulate the overall  solution  density  and
compressibility.
 The excess density and excess compressibility, with respect to an
\begin{figure}
\begin{center}
\epsfig{file=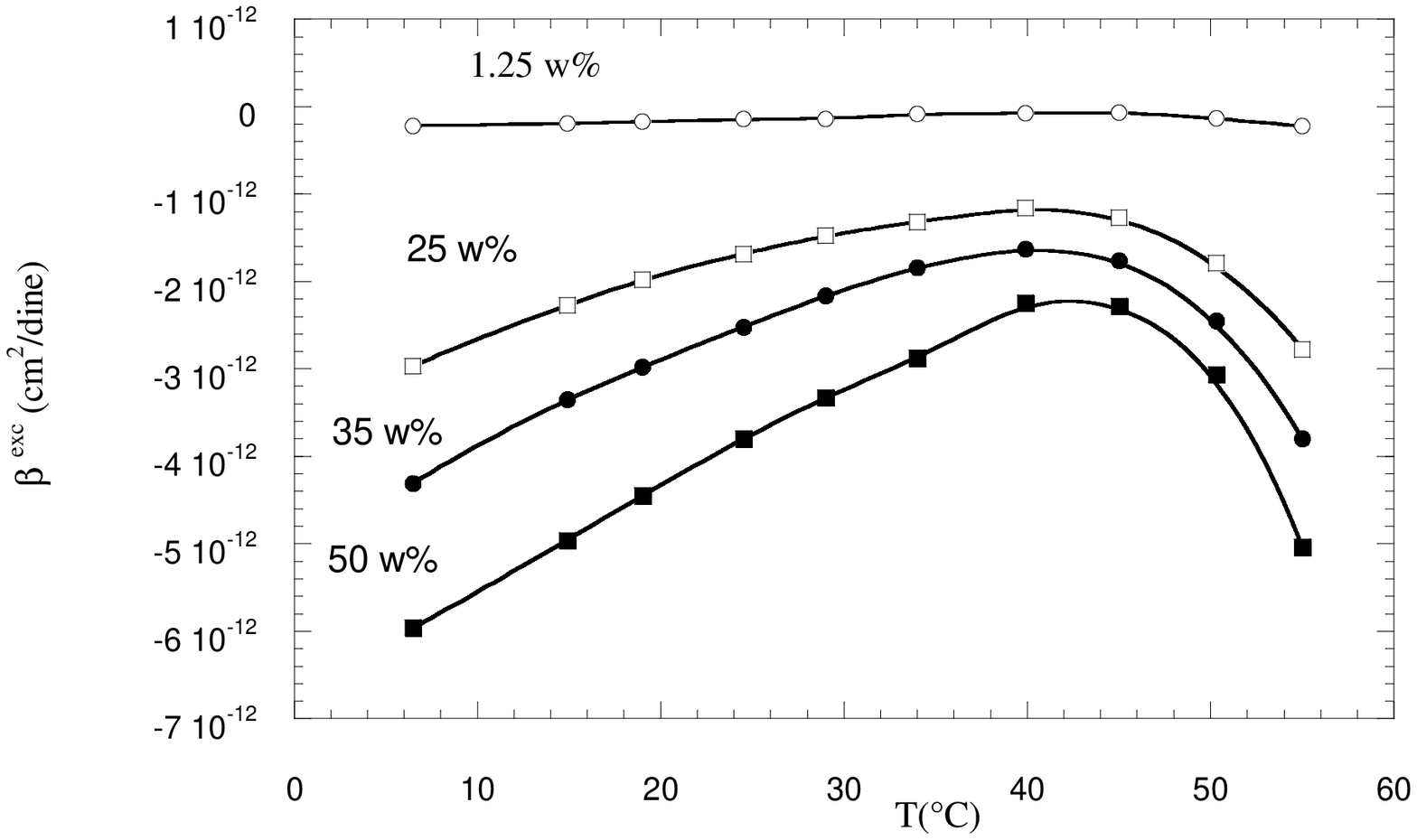,height=10cm}
\caption{
The  Excess  compressibility  as  a  function  of
temperature    and    at    four   different    surfactant
concentrations are presented.}
\label{fig:betaexvst}
\end{center}
\end{figure}
\begin{figure}
\begin{center}
\epsfig{file=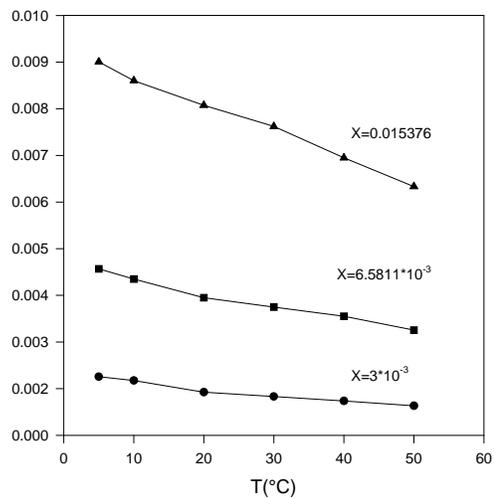,height=8cm}
\caption{
The excess density with respect to an ideal systems
of  surfactant  and water are reported at three  different
amphiphilic concentrations.}
\label{fig:dedid}
\end{center}
\end{figure}
ideal solution of surfactant and water, are reported in Fig.
\ref{fig:dedid}  and Fig. \ref{fig:betaexvst} respectively.
The density results concern dilute surfactant
solution  (2 $ w\%$),  in  this  case two different  regimes  can  be
observed  above  and  below about $20^{\circ}C$. At  this  temperature  is
generally  indicated the presence of the sphere to rod  transition
in $C_{12}E_6$-water solutions \cite{brown}, from a theoretical point
of  view the area per polar head corrispondently strongly reduces
\cite{puvv}.
Again this evidence supports the role of the interface in this
shape transition. For compressibility the available concentrations
are too high to present any sphere to rod transition \cite{arribri}.

\section{Discussion and Conclusions}
\begin{figure}
\begin{center}
\epsfig{file=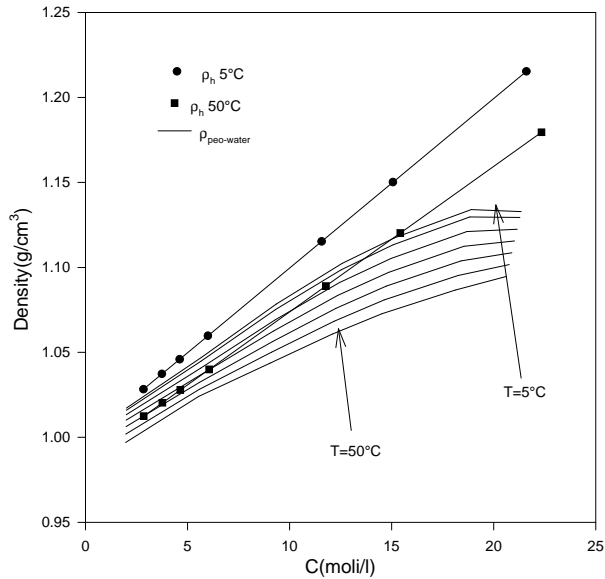,height=8cm}
\caption{
Comparison of the interfacial density  at  5  and
$50^{\circ}C$,  obtained as a function of the  interfacial
volume,  is  compared with the experimental dependence  on
concentration   of   EO-water   solutions   at   different
temperatures.}
\label{fig:ae6}
\end{center}
\end{figure}

 From  density  and  compressibility it appears  that  the
liquid phase of pure $C_{12}E_{6}$ is characterized by two
separated region: the oil and the EO phases; in this  case
the  EO region exactly resemble the pure EO polymer liquid
phase.  Our results of density and compressibility at  the
micellar  interface are parameterized with the interfacial
volume, it means that we have a set of different value for
the  interfacial properties which depend on the  local  EO
molarity.  In  a similar manner as in the pure  surfactant
case,  we  can compare the density dependence  on  the  EO
molarity at the micellar interface with the density of EO-
water   solutions   as  a  function  of   the   EO   molar
concentrations. The results are reported in Fig. \ref{fig:ae6}.
It is
evident  that  there could exist many  solutions  for  the
interface  compatible  with different  conditions  of  the
reference mixture, a priori all of them could be accepted.
A  similar procedure could not be performed in the case of
compressibility since the quantities $N^{h}_{w}$, i.e. the
number of the hydrated water molecules per monomer, is not
easily eliminated in eq. \ref{eq:comprh} as for density
(see eq. \ref{eq:rhoh}  and \ref{eq:rhoh2}).
 An independent determination of the volume $V_h$ has been
obtained  comparing  the interfacial  dielectric  constant
with  the  one of EO-water solutions. Since the dielectric
constant  is  related  to the local polarizability,  which
mainly  depends  on  the  degree of  connectivity  of  the
hydrogen  bond network, then the $V_h$ obtained with  this
technique  represent reliable value. Accordingly  we  used
such  $V_h$  values in order to fix from the  density  the
number  of hydrated water molecules. Using this number  in
the  eq.  \ref{eq:comprh} for the solution compressibility we  can  get
explicit values for the interfacial compressibility. As  a
results we obtained values with are higher than those in EO-water
solutions, and are very sensitive to the variation of  the
hydration number.
 It is interesting to notice that the coincidence of the $\rho$
and  $\beta$  behavior  as a function of temperature,  i.e.  the
existence  of  a  common temperature  at  which  both  the
quantity  are  independent  on  concentration,  could   be
indicative  of  a  condition at which  the  second  Virial
coefficient reduce to zero.
 The  shape change in nonionic micellar solution, from the
sphere to rod, is evident on the excess density where  two
different   regimes  are  present  above  and  below   the
transition  temperature. These two regimes, in our  model,
reflect two different conformation and degree of hydration
of  the  micellar interface. The change of the interfacial
properties  with temperature is then a complex combination
of  change  in  the EO conformation and of the  degree  of
hydration and cannot be reduced to excluded area per polar
\cite{puvv}.

\end{document}